\documentclass[prd,aps,floats,twocolumn,showpacs,preprintnumbers,floatfix]{revtex4}

\usepackage{epsfig}

\begin{document}

\preprint{CERN-PH-TH/2005-086}

\title{
Large Scale Structure in Bekenstein's theory of relativistic Modified Newtonian Dynamics} 
\author{C.~Skordis$^1$,  D.~F.~Mota$^2$,  P.~G.~Ferreira$^{1,3}$, C.~B\oe hm$^{4,5}$}
\affiliation{$^1$Astrophysics, University of Oxford, DWB, Keble Road, Oxford OX1 3RH, United Kingdom \\
$^2$Institute for Theoretical Astrophysics, University of Oslo, N-0315 Oslo, Norway\\
$^3$African Institute for Mathematical Sciences (AIMS), 6-8 Melrose Road, Muizenberg 7945, South Africa\\
$^4$TH Division, PH Department, CERN 1211, Geneve 23, Switzerland\\
$^5$LAPTH, UMR 5108, 9 chemin de Bellevue -BP 110, 74941 Annecy-Le-Vieux, France
}

\begin{abstract}
A relativistic theory of modified gravity has been recently proposed by Bekenstein.
The tensor field in Einstein's theory
of gravity is replaced by a scalar, a vector, and a tensor field which
interact in such a way to give Modified Newtonian Dynamics (MOND)
in the weak-field non-relativistic limit.
We study the evolution of the universe in such
a theory, identifying its key properties and comparing it with the standard
cosmology obtained in Einstein gravity.
The evolution of the scalar field is akin to
that of tracker quintessence fields. We expand the theory
to linear order to find the evolution of perturbations on large scales. The
impact on galaxy distributions and the cosmic microwave background is
calculated in detail.  We show that it may be possible 
to reproduce observations
of the cosmic microwave background and galaxy distributions with Bekenstein's
theory of MOND.
\end{abstract}

\pacs{98.90.Cq, 98.65.Dx, 98.70.Vc} 

\maketitle

\renewcommand{\thefootnote}{\arabic{footnote}} \setcounter{footnote}{0}
 The current
model of the Universe is based on a few simple assumptions and can
explain a multitude of observation. Yet, to be able to explain the
structure of galaxies and clusters of galaxies, it is essential to
postulate the existence of some invisible substance, called dark
matter. Although there are reports of tentative discoveries of dark
matter \cite{bhs2005},  there is no proven theory
or direct observation of a dark matter particle as yet. A less
explored route is that our current theory of gravity might be
incomplete. Given that the existence of dark matter is inferred from
its gravitational effects on the dynamics of astrophysical bodies,
it may be that the theory used to link the dynamics to the
mass is incorrect. Milgrom has proposed a modification of
Newtonian dynamics, known as MOND \cite{M1983}. In MOND, Newton's
second law in a gravitational field is modified to $\mu(|{\vec
a}|/a_0){\vec a}=-\nabla \Phi$ where ${\vec a}$ is the acceleration,
$\Phi$ is the Newtonian potential and $\mu(x)$ is a function with a
scale $x\simeq 1$. For $|{\vec a}|>a_0$ we have that $\mu(x)=1$. With
$|{\vec a}|<a_0$ we have that $\mu(x)\simeq x$. Clearly for small
accelerations, Newtonian theory is no longer valid.
 Milgrom's theory has been extremely successful in
explaining a number
of observational properties of galaxies~\cite{galaxy_MOND}.
It has suffered from a fatal flaw in that it is
not generally covariant and hence cannot be studied in a general setting.

Bekenstein has recently solved this problem \cite{B2004}.
Building on a series of developments \cite{PreTeVeS},
he has proposed a generally covariant theory which has in
the nonrelativistic, weak-field limit, Milgrom's modified theory.
Bekenstein's theory
has two metrics. One of the metrics,
${\tilde g}_{\mu\nu}$ has its dynamics governed
by the Einstein-Hilbert action,
\begin{eqnarray}
 S_g=\frac{1}{16\pi G}\int d^4x\sqrt{-{\tilde g}}{\tilde R}, \nonumber
\end{eqnarray}
 where $G$ is Newton's constant and ${\tilde R}$ is the scalar curvature
of ${\tilde g}_{\mu\nu}$. We shall call the frame of this metric the
``Einstein Frame'' (EF). The second metric, $ g_{\mu\nu}$
 is minimally coupled to all the matter fields in the Universe.
We shall call the frame of this metric the ``Matter Frame'' (MF).
All geodesics
are calculated in terms of this second metric. The two metrics are related
through
${g}_{\mu\nu}=e^{-2\phi}({\tilde g}_{\mu\nu}+A_\mu A_\nu)-e^{2\phi}
A_\mu A_\nu$.
Two fields are required to connect the two metrics. The scalar field, $\phi$
has  dynamics given by the action
\begin{eqnarray}
S_s=-\frac{1}{16\pi G}\int d^4x  \sqrt{-{\tilde g}}
  \left[\mu\;\left({\tilde g}^{\mu\nu} - A^{\mu}A^{\nu}\right)\phi_{,\mu}\phi_{,\nu} +V(\mu)\right],\nonumber
\end{eqnarray}
where $\mu$ is a nondynamical field and $V$ is
a free function which can be chosen to give the correct nonrelativistic
MOND limit and depends on two free parameters, $\ell_B$ and $\mu_0$ (related to
$\kappa$ in \cite{B2004}  as $\mu_0 = 8\pi/\kappa$). The unit timelike
vector field, $A_{\mu}$ has dynamics given by the action
\begin{eqnarray}
S_v= -\frac{1}{32\pi G}\int d^4x \sqrt{-{\tilde g}} \bigg[  K F^{\alpha\beta}
F_{\alpha\beta}  -2\lambda
(A^\mu A_\mu+1) \bigg], \nonumber
\end{eqnarray}
where $F_{\mu\nu} = A_{\mu,\nu} - A_{\nu,\mu}$,
indices are raised with ${\tilde g}$
and where $K$ is the third parameter in this theory. The Lagrange multiplier $\lambda$
 is completely fixed by variation of the action.

We wish to study the evolution of a homogeneous and isotropic
universe in such a theory.  Observers are defined in the MF where the 
 line element is $ds^2=a^2(-d\eta^2+dr^2)$. The scale factor 
is related to the metric in the EF,
$d\tilde{s}^2=b^2(-e^{-4\phi}d\eta^2+dr^2)$ through $a=be^{-\phi}$. 
The modified Friedmann equation becomes
\begin{eqnarray}
3\frac{\dot b^2}{b^2}=a^2\left[\frac{1}{2}e^{-2\phi}(\mu V'+V)
+8\pi G e^{-4\phi}{\rho}\right], \nonumber
\end{eqnarray}
where $\mu$  can be found by inverting
${\dot \phi}^2=\frac{1}{2}a^2e^{-2\phi}\frac{dV}{d\mu}$
and the energy density $\rho$ does not include $\phi$. Homogeneity
and isotropy and the constraint in the action  imply that
$A_\mu$ is fixed as $A_\mu= a e^{-\phi}(1,0,0,0)$.
The background dynamics is complete with an equation for $\phi$:
\begin{eqnarray}
{\ddot \phi} &=& -a^2e^{-2\phi}V'-\frac{1}{U}\bigg[2(\mu-\frac{V'}{V''})
\frac{\dot b}{b}{\dot \phi} \nonumber \\
 && + 4\pi G a^2 e^{-4\phi}(\rho+3P)\bigg], \nonumber
\end{eqnarray}
where $U=\mu+2V'/V''$ and, once again, $P$ does not include the pressure
from the scalar field $\phi$.

A choice of $V$ will pick out a given theory. As a first guess,
Bekenstein has proposed
\begin{eqnarray}
V=\frac{3\mu_0^2}{128\pi\ell_B^2 }
 \left[\hat{\mu}(4+2\hat{\mu}- 4\hat{\mu}^2+\hat{\mu}^3) +2\ln(\hat{\mu}-1)^2\right],
\nonumber 
\end{eqnarray}
where $\hat{\mu} = \mu/\mu_0$.
This potential will lead to the prescription proposed by Milgrom in
the nonrelativistic region. The evolution of this coupled set of
equations is essentially
 insensitive to $\ell_B$ and independent of $K$ but
 is well determined in terms of $\mu_0$. If we
define the physical Hubble parameter $H$, we can rewrite
 the modified Friedmann equations in the form
$3H^2 = 8\pi G_{eff}(\rho+\rho_\phi)$
where the effective Newton's constant has the form
$G_{eff}=G{e^{-4\phi}}/({1+\frac{d\phi}{d\ln a}})^2$
and the energy density in $\phi$ is
\begin{eqnarray}
\rho_\phi=\frac{e^{2\phi}}{16\pi G}\left( \mu V'+V \right).
\nonumber
\end{eqnarray}
This system exhibits the tracking behavior witnessed in some scalar
field theories of quintessence \cite{Scaling}. In particular one
finds that  for a wide range of initial conditions, $\phi$ evolves
to a slowly varying function of time and the
relative energy density in $\phi$ reaches an attractor solution of the form
$\Omega_\phi =3/(2\mu_0)$ in the radiation era and
$\Omega_\phi = 1/(6\mu_0)$ in the matter and $\Lambda$
 eras.

In Fig. 1 we solve the equations numerically for a low
value of  $\mu_0$ to illustrate the tracking behavior of $\phi$.
\begin{figure}[ht]
\epsfig{file=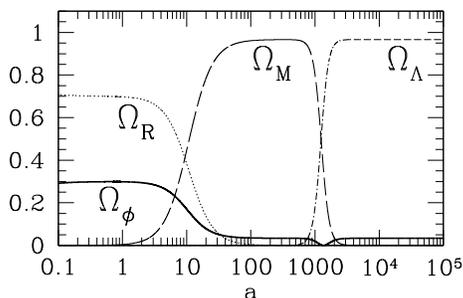,width=7.3cm,height=7.3cm}
 \vskip -1.28in
\caption{
The relative energy densities in $\phi$ (thick solid
line), radiation (dotted line), matter (dashed line) and $\Lambda$
(dot-dashed line) for $\mu_0=5$ as a function of the scale
factor ($a$ is in arbitrary units). Note that the energy density in the
scalar field tracks the dominant form of energy at each instance in
time.   \label{Fig1}}
\end{figure} 
As with other tracking systems we can constrain the energy
density in $\phi$  at nucleosynthesis \cite{BBNbounds}.
The abundance
of light elements is extremely sensitive to
the  expansion rate at 1 MeV and can, for example, be used
to constrain the number of relativistic degrees of freedom
at that time. Recent measurements of the $^4$He mass fraction and
the deuterium abundance  leads
to a bound at energies
of $1$ MeV of $\Omega_\phi<0.045$
at the 95$\%$ confidence level. This leads to $\mu_0>33$ and hence
 the scalar field will make up less than $0.5\%$ of
the total energy density during the matter and $\Lambda$ dominated
eras.

We now turn to evolution of linear perturbations on this background. This will
allow us to link this theory with observations of galaxy clustering on large
scales as well as with the anisotropies in the Cosmic Microwave Background (CMB).
 The main problem that a MOND theory containing only baryons has to
confront is the damping of perturbations during the recombination
era. Indeed in a pure baryonic universe evolving under Einstein gravity,
the weak coupling of baryons and photons during the recombination era will
lead to Silk damping, the collisional propagation
 of radiation from overdense to underdense regions \cite{BDM}.
 In the standard adiabatic model with just baryons, the
matter power spectrum is severely suppressed on galactic scales. If
MOND is to succeed, it must overcome the Silk damping on these
scales.

There are
three fields to perturb around the background (we identify the homogeneous
part of a quantity $X$ by ${\bar X}$).  The MF metric perturbations
are given in the Conformal Newtonian gauge by ${ g}_{00}=-a^2(1+2{ \Psi})$ and
${g}_{ij}=a^2(1-2{ \Phi})\delta_{ij}$.  The scalar field perturbation is
given by
$\phi={\bar \phi}+\varphi$. The vector field perturbation is defined by
$A_\mu=ae^{-\bar{\phi}}(\bar{A}_\mu+\alpha_\mu)$ where  the scalar components,
$\alpha$ and $E$, of $\alpha_\mu=( \Psi-\varphi, {\vec \alpha})$
are given by $\nabla^2\alpha\equiv \nabla\cdot{\vec \alpha}$
and $\nabla^2E\equiv \nabla\cdot {\vec E}$ where we use the field strength
tensor of $A_\mu$, $F_{\mu\nu}$ to define the ``electric field'' through
$E_i= {\bar A}^\mu F_{i\mu}$.
 Similarly, in EF, we have
${\tilde g}_{00}=-b^2e^{-4\bar{\phi}}(1+2\tilde{\Psi})$,
 ${\tilde g}_{0i}=-b^2\tilde{\zeta}_{,i}$
and ${\tilde g}_{ij}=b^2(1-2{ \Phi})\delta_{ij}$ which give 
${\tilde \Psi}=\Psi-\varphi$, ${\tilde \Phi}=\Phi-\varphi$ and 
$\tilde{\zeta} = -(1-e^{-4\bar{\phi}})\alpha$.

The evolution equations for the matter fluid remain unaltered if
expressed in terms of the MF variables. That is if we expand densities
as $\rho={\bar \rho}(1+\delta)$ and use the standard definition for
momentum of the fluid, $\nabla^2\theta=\nabla\cdot{\vec v}$, the evolution
equations remain the same as in Einstein gravity. Two new sets of
evolution equations must be introduced. For the scalar field
perturbations, we have
\begin{eqnarray}
{\dot \varphi}&=&-\frac{ae^{-{\bar \phi}}}{2U}\gamma+{\dot {\bar \phi}}
{\tilde \Psi} \nonumber \\
{\dot \gamma}&=&-3\frac{\dot b}{b}\gamma+\frac{\bar \mu}{a}e^{-3{\bar \phi}}
k^2(\varphi+{\dot {\bar \phi}}\alpha)-2\frac{e^{\bar \phi}}{a}
\left(3{\dot {\tilde \Phi}} + k^2\tilde{\zeta}\right) 
\nonumber \\ & &+ 8\pi G ae^{-3{\bar \phi}}{\bar \rho}
[(1+3c^2_s)\delta+(1+3w)({\tilde \Psi}-2\varphi)] \nonumber
\end{eqnarray}
and for the vector field we have
\begin{eqnarray}
{\dot \alpha}&=&\frac{e^{\bar \phi}}{a}E+{\tilde \Psi}+
({\dot {\bar \phi}}-\frac{\dot a}{a})\alpha \nonumber \\
K\frac{e^{2{\bar \phi}}}{a^2}({\dot E}+2{\dot {\bar \phi}}E)&=&
-\mu\frac{e^{\bar \phi}}{a}{\dot {\bar \phi}}\left(\varphi 
 - \dot{\bar{\phi}}\alpha\right) +16\pi Gae^{-{\bar \phi}} \nonumber \\
& & \times 
\sinh(2{\bar \phi})(1+w){\bar \rho}(\theta-\alpha)  \nonumber
\end{eqnarray}
where $w=\bar{P}/\bar{\rho}$ and $c_s^2=\delta P/\delta \rho$.
The perturbed Einstein equations allow us to identify the
gravitational potentials through:
\begin{eqnarray}
2k^2{\tilde \Phi}&=&-2e^{4{\bar \phi}}\frac{\dot b}{b}k^2{\tilde \zeta}
-e^{4{\bar \phi}}{\dot {\bar \phi}}\left\{-ae^{-{\bar \phi}}\gamma
+6{\bar \mu}\frac{\dot b}{b}\varphi\right\}\nonumber \\ & &
-8\pi G a^2{\bar \rho}\left\{\delta+3(1+w)\frac{\dot b}{b}\theta
-2\varphi\right\}-Kk^2\frac{e^{\bar \phi}}{a}E, \nonumber \\
{\dot {\tilde \Phi}}&=&4\pi G a^2e^{-4{\bar \phi}}(1+w){\bar \rho}\theta
+{\bar \mu}{\dot {\bar \phi}}\varphi-\frac{\dot b}{b}{\tilde \Psi},\nonumber
\\ 
 \tilde{\Psi}  &=& \tilde{\Phi} + e^{4\bar{\phi}}\left[\dot{\tilde{\zeta}}
+ 2\left(\frac{\dot{b}}{b} + \dot{\bar{\phi}}\right) \tilde{\zeta} \right]
  - \frac{12\pi G}{k^2} a^2 (1+w) \rho \sigma, 
\nonumber 
\end{eqnarray}
where $\sigma$ is the total shear from matter fluids.

We have modified CMBEASY, a publicly available numerical Einstein-Boltzmann
solver to incorporate the modified background and perturbation equations
\cite{CMBEASY}. The evolution equations have been implemented in
both the conformal Newtonian and synchronous gauge to check for
consistency~\cite{S2005}. We have restricted ourselves to a flat Universe with
a cosmological constant but considered the possibility of massive neutrinos.

\begin{figure}[ht]
\epsfig{file=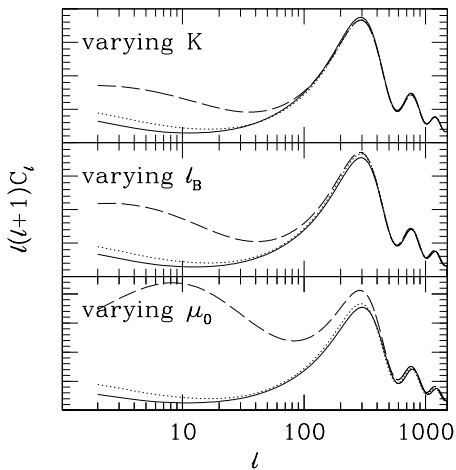,width=7.3cm,height=7.3cm}
\vskip -0.37in
\caption{
The effect of the MOND parameters on the power of spectrum
of the CMB. Top panel: $\mu_0=200$, $\ell_B=100$Mpc and
$K=1$ (solid), $0.1$ (dotted) and $0.08$ (dashed);
 Middle panel: $\mu_0=200$, $K=0.1$ and
$\ell_B=1000$Mpc (solid), $100$Mpc (dotted) and $10$Mpc (dashed);
 Bottom panel: $K=0.1$, $\ell_B=100$Mpc and
$\mu_0=1000$ (solid), $200$ (dotted) and $150$ 
 (dashed).\label{Fig2}}
\end{figure}

The parameters $\mu_0$, $K$ and $\ell_B$ may introduce major
modifications in the morphology of perturbations. A low $\mu_0$, low $\ell_B$ and
low $K$ will lead to a change in the  growth rate.
As we can see in Fig. 2, the effect is to introduce
an integrated Sachs Wolfe term which can be quite significant. 
 For example, for sufficiently small $\ell_B$, the structure of the
angular power spectrum of the CMB can be completely modified with 
an excess of large scale power overwhelming structure on the smallest
scales.  We can see the effect of modifying $\ell_B$ in the lower
panel of Fig. 2.
 Clearly, the CMB can place quite stringent
constraints on the values of these parameters.
A further possible effect, that we have not included here, is the effect
of secondary anisotropies such as the Sunyaev-Zel'dovich or Ostriker-Vishniac
effects, which may leave a different signature than the standard cold dark
matter $\Lambda$CDM model.

Since the baryon content is set by the abundance of light elements,
we must compensate with a high value of the cosmological constant,
i.e. with $\Omega_\Lambda\simeq 0.95$. An obvious consequence of
this is that the angular-distance relation will be modified as
compared to the standard adiabatic $\Lambda$CDM universe
\cite{PeakOmega}. Indeed the position of the peaks in the angular
power spectrum of the CMB will be shifted to higher $l$s which would lead
to a severe mismatch with the current available data from the Wilkinson Microwave 
Anisotropy Probe and other experiments. A natural solution to this is to include a small
component of massive neutrinos, $\Omega_\nu\simeq0.15$. As we can
see in the top panel of Fig. 4, with this modification we can
reproduce the temperature anisotropy data.

\begin{figure}[ht]
\epsfig{file=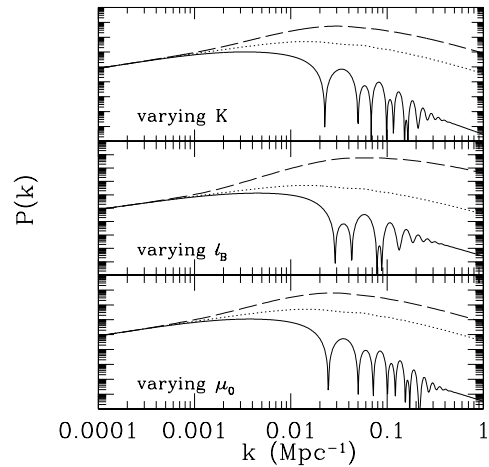,width=7.3cm,height=7.3cm}
\vskip -0.37in
\caption{
The effect of the MOND parameters on the power of spectrum
of the baryonic density fluctuations.
 Top panel: $\mu_0=200$, $\ell_B=100$Mpc and
$K=1$ (solid), $0.1$ (dotted) and $0.08$ (dashed);
 Middle panel: $\mu_0=200$, $K=0.1$ and
$\ell_B=1000$Mpc (solid), $100$Mpc (dotted) and $10$Mpc (dashed);
 Bottom panel: $K=0.1$, $\ell_B=100$Mpc  and
$\mu_0=1000$ (solid), $200$ (dotted) and $150$
 (dashed).\label{Fig3}}
\end{figure}
The main question we have raised is whether MOND dynamics can inhibit
the damping of small scale perturbations in the coupled
baryon-photon fluid during recombination. Recall that in the
adiabatic CDM model, perturbations in the dark matter, $\delta_{C}$,
are undamped during recombination. The Newtonian potential, which is
roughly given by $k^2\Phi\simeq 4\pi G(\rho_B\delta_B
+\rho_C\delta_C)$ will not be erased if $\rho_C$ is sufficiently
large, even though $\delta_B\rightarrow 0$ through recombination
\cite{P81}. In the MOND universe we find an analogous effect; we
now have $k^2\Phi\simeq 4\pi G\rho_B(\delta_B-2\varphi)$. The
perturbation in the scalar field will support the perturbations
through recombination yet still allow the damping of anisotropies
in the photon fluid.
 Unlike the case of dark matter however, the
coupling between the scalar field and the metric is such that
$\rho_\phi$ does not play a role in the magnitude of the effect.
Even for minute values of $\Omega_\phi$ we can still have a
non-negligible effect. As we can see in Fig. 3, 
the net result is that decreasing $\mu_0$, $\ell_B$ or
$K$ will boost small scale power in such a way as to overcome the
damping of perturbations.
\begin{figure}[ht]
\epsfig{file=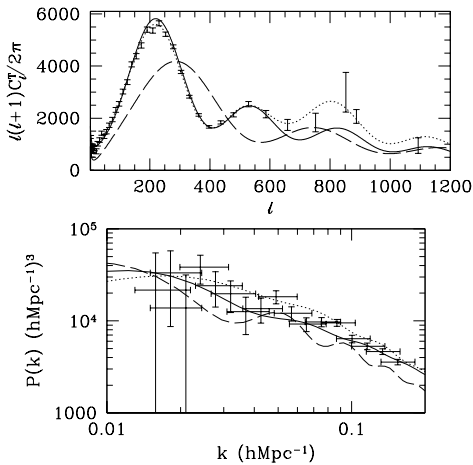,width=7.3cm,height=7.3cm} 
\vskip -0.37in
\caption{
The angular power spectrum of the CMB (top panel) and the
power spectrum of the baryon density (bottom panel) for a MOND
universe (with $a_0\simeq 4.2\times10^{-8}cm/s^2)$ with $\Omega_\Lambda=0.78$ and
$\Omega_\nu = 0.17$ and $\Omega_B=0.05$ (solid line), for a MOND universe
$\Omega_\Lambda=0.95$ and $\Omega_B=0.05$ (dashed line)
 and for the $\Lambda$-CDM model (dotted line).
A collection of data points from CMB experiments and Sloan are overplotted.
\label{Fig4}}
\end{figure}
  This is an intriguing effect that goes in
tandem with what we saw in the CMB. While decreasing $\ell_B$ (and a
sufficiently small $K$ and $\mu_0$) will contaminate the large scale
power in the angular power spectrum of the CMB, it can also play a
role in counteracting Silk damping of density perturbations.

Given these two effects on the dynamics of
large scale structure, is it possible to construct a MOND universe which
can reproduce current observations of the CMB and galaxy surveys? There is
clearly a competition 
between overproducing large scale power in the CMB but
also overcoming damping on small scale. In Fig. 4
we present two MOND universes compared to data \cite{WMAP,SDSS}.
As mentioned above, a universe with
a very large contribution of $\Lambda$ will not fit the current CMB data. By
having the three neutrinos with a mass of $m_\nu\simeq 2$ eV each
we are able to resolve this mismatch.
 With an appropriate choice of $K$, $\mu_0$ and $\ell_B$
it is possible to reproduce the power spectrum of galaxies as inferred from
the Sloan Digital Sky Survey \cite{SDSS}. 
The possibility of using massive
neutrinos to resolve some of the problems with clusters in a MOND
universe has been mooted in \cite{S2003}.

 We have focused on one very
specific model proposed by Bekenstein with a somewhat artificial
potential for the new degrees of freedom. This phenomenological
approach needs a firmer theoretical underpinning which might come
from the various approaches which are being taken in the context of
brane worlds, M-theory and a rich array of theories of modified
gravity. However, Bekenstein's theory can play an important role in
opening up an altogether different approach to the dark matter
problem. It serves as a proof of concept which will clearly lead to
a new, very different view of the role played by the gravitational
field in cosmology.

{\it Acknowledgments}:
We thank J. Bekenstein, J. Binney, M. Doran, J. Dunkley, O. Elgaroy, J-M Frere, D. Hooper, 
S. Pascoli and O. Vives 
for discussions. C.S is supported by PPARC
Grant No. PPA/G/O/2001/00016. D.F.M. is supported by Research 
Council of Norway through Project No.  159637/V30.



\begin{thebibliography}{99}

\bibitem{bhs2005} G. Bertone, D. Hooper and J. Silk,
Phys. Rep. {\bf 405}, 279 (2004).
\bibitem{M1983} M. Milgrom, Astroph. J. {\bf 270}, 365 (1983); {\bf 270}, 371 (1983);
{\bf 270}, 384 (1983).
\bibitem{galaxy_MOND} R.~Sanders and S.~McGaugh, Annu. Rev. Astron. Astrophys. {\bf 40}
263 (2002); S.~McGaugh and E. de Blok, Astrophys. J. {\bf 499}, 66 (1998).
\bibitem{B2004}J.D. Bekenstein, Phys. Rev. D {\bf 70}, 083509 (2004).
\bibitem{PreTeVeS} J.~D.~Bekenstein and M.~Milgrom, Astrophys. J. {\bf 286}, 7 (1984);
J.D. Bekenstein, Phys. Lett. B {\bf 202}, 497 (1988);
R.H.Sanders, Astrophys. J. {\bf 480}, 492 (1997).
\bibitem{Scaling} B.~Ratra and P.~J.~Peebles, Phys. Rev. D {\bf 37}, 3406 (1988);
C.~Wetterich, Nucl. Phys. {\bf B252}, 302 (1988); E.~Copeland {\it et al.},
Ann. N.Y. Acad. Sci. {\bf 688}, 647 (1993); P.G.Ferreira and M.~Joyce, Phys.
Rev. Lett. {\bf 79}, 4740 (1997); C.~Skordis and A.~Albrecht, Phys. Rev. D{\bf 66}, 043523 (2002).
\bibitem{BBNbounds}R.~Bean, S.~Hansenand  A. Melchiorri, Phys. Rev. D{\bf 64},
103508 (2001).
\bibitem{BDM} P.J.Peebles and J.T.Yu Astrophys. J. {\bf 162}, 815 (1970); M.L.Wilson
and J.Silk, Astrophys. J. {\bf 243}, 14 (1981); L.~Griffiths, A.~Melchiorri and J.~Silk,
Astrophys. J. {\bf 553}, L5 (2001).
\bibitem{CMBEASY} M.Doran,  {\tt http://www.arxiv.org/} astro-ph/0302138.
\bibitem{S2005}C. Skordis, astro-ph/0511591.
\bibitem{PeakOmega} G.Efstathiou and D.Bond, Mon.Not.R.Astron.Soc. {\bf 304}, 75 (1999).
\bibitem{P81}P.J.Peebles, Astrophys. J. {\bf 248}, 885 (1981).
\bibitem{WMAP}C. Bennett {\it et al.} (WMAP collaboration), Astrophys. J. Supp. {\bf 148}, 1 (2003);
X.~Wang {\it et al.}, Phys.~Rev.~D {\bf 68}, 123001 (2003);
C.~L.~Kuo, {\it et al.}, Astrophys. J. {\bf 600}, 32 (2004);
J.~Ruhl {\it et al.}, Astrophys. J, {\bf 599}, 786 (2003);
T.J~Pearson {\it et al.}, Astrophys.~J. {\bf 591}, 556 (2003);
P.F.~Scott {\it et al.}, Mon.~Not.~R.~Astron.~Soc. {\bf 341}, 1066 (2003).
\bibitem{SDSS} M.~Tegmark {\it et al.}, Astrophys. J. {\bf 606}, 702 (2004).
\bibitem{S2003}R. Sanders,  Mon.~Not.~R.~Astron.~Soc. {\bf 342}, 901 (2003).
\end{thebibliography}
\end{document}